\documentclass[12pt]{article}
\usepackage{epsfig}
\usepackage[english]{babel}
\usepackage[T1]{fontenc}
\usepackage{graphicx}
\usepackage{amsmath, amssymb, latexsym, amsopn}
\usepackage{mathtools}
\usepackage{authblk}
\usepackage{hyperref}
\usepackage{natbib}

\newcommand{\av}[1]{\langle{#1}\rangle}

\title{Emergence of spatial curvature}

\author[1]{Krzysztof Bolejko\thanks{krzysztof.bolejko@sydney.edu.au}}
\affil[1]{Sydney Institute for Astronomy, School of Physics, A28, The University of Sydney, NSW, 2006, Australia
}

\begin{document}

\maketitle

\begin{abstract}
This paper investigates the phenomenon of emergence of spatial curvature.
This phenomenon is absent in the Standard Cosmological Model, which has a flat and fixed spatial curvature (small perturbations are considered in the Standard Cosmological Model but their global average vanishes, leading to spatial flatness at all times).
This paper shows that with the nonlinear growth of cosmic structures the global average deviates from zero.  The analysis is based on the {\em silent universes} (a wide class of 
inhomogeneous cosmological solutions of the Einstein equations). The initial conditions are set in the early universe as perturbations around the $\Lambda$CDM model with $\Omega_m = 0.31$, $\Omega_\Lambda = 0.69$, and $H_0 = 67.8$ km s$^{-1}$ Mpc$^{-1}$. As the growth of structures becomes nonlinear, the model deviates from the  $\Lambda$CDM model, and at the present instant if averaged over a domain ${\cal D}$ with volume $V = (2150\,{\rm Mpc})^3$ (at these scales the cosmic variance is negligibly small) gives: $\Omega_m^{\cal D} = 0.22$, $\Omega_\Lambda^{\cal D}  = 0.61$, $\Omega_{\cal R}^{\cal D}  = 0.15$ (in the FLRW limit $\Omega_{\cal R}^{\cal D} \to \Omega_k$), and $\av{H}_{\cal D} = 72.2$ km s$^{-1}$ Mpc$^{-1}$. Given the fact that low-redshift observations favor higher values of the Hubble constant and lower values of matter density, compared to the CMB constraints, the emergence of the spatial curvature in the low-redshift universe could be a possible solution to these discrepancies.
\end{abstract}

\section{Introduction}

Astronomical constraints on the spatial curvature are often derived, not by a direct measurement, but by 
fitting a spatially homogeneous and isotropic model, or a linearly perturbed FLRW model, to observational data.
The tightest constraints come from the early universe: the CMB data (TT+LowP+lensing) constrain the curvature at a percent level. If accompanied with the BAO data then
the spatial flatness is confirmed with a  sub-percent precision \citep{2016A&A...594A..13P}.
These constraints fit nicely with inflationary scenarios that predict spatial flatness of the early universe \citep{1981PhRvD..23..347G}.

However, the spatial curvature of the FLRW models is very rigid, so fitting these models to the data is not equivalent to direct measurements, which do not provide such tight constraints \citep{2015PhRvL.115j1301R}. Investigations of nonlinear dynamics 
by \citet{2008CQGra..25s5001B} and also \citet{2011CQGra..28p5004R} suggest that the overall spatial curvature of our universe may in fact be negative. Recently, using a Styrofoam model that consisted of the Szekeres cells, it was shown how the spatial curvature emerges due to the evolution of the cosmic structures \citep{2017arXiv170402810B}. 
This paper investigates this phenomenon  further using a more general class of {\em silent universes}. The investigation is based on the Monte Carlo simulation that consist of 
$2 \times 10^6$ worldlines. These worldlines are evolved from the early universe to the present instant. Each worldline is characterized by density, expansion rate, shear, and Weyl curvature.
From these quantities, the spatial curvature is evaluated, and it is shown that the global average evolves from zero and reaches non-negligible, negative values at the present instant. The structure of this paper is as follows: Sec. \ref{silent} briefly presents the {\em silent universes};
Sec. \ref{emcur} presents the results of the Monte Carlo simulation and shows how the spatial curvature emerges due to nonlinear evolution (in the linear regime
 negative curvature of voids is compensated by positive curvature of overdense regions and so the global spatial curvature is zero); Sec \ref{conclusion} discusses the obtain results and speculates on possibility of detecting this phenomenon.

\section{ Silent Cosmology} \label{silent}

The evolution of a relativistic system is set by its content ($T_{ab}$) and the space-time geometry ($G_{ab} - \Lambda g_{ab}$).
In the 3+1 split, the system can be reduced to scalars (density, expansion rate), vectors (particle acceleration, rotation, particle flux, energy transfer, entropy flux), and tensors (shear, anisotropic stress-tensor, magnetic and electric parts of the Weyl curvature). On scales, above 2-5 Mpc (cf. matter horizon defined by \citet{2009MNRAS.398.1527E}) contribution from particle flux, pressure and viscosity, and rotation can be neglected, and the solution of the Einstein equations reduces to the {\em silent universe}, where each worldline evolves independently (hence `silent'), and the whole system is described by 4 scalars: density $\rho$, expansion $\Theta$, shear $\Sigma$, and the Weyl curvature ${\cal W}$ \citep{1995ApJ...445..958B,1997CQGra..14.1151V}

\begin{eqnarray}
 \dot \rho &=& -\rho\,\Theta, \label{PFev1}\\
\dot \Theta &=& -\frac{1}{3}\Theta^2-\frac{1}{2}\,\kappa \rho-6\,\Sigma^2 + \Lambda,\label{PFev2}\\
\dot \Sigma &=& -\frac{2}{3}\Theta\,\Sigma+\Sigma^2-{\cal W},\label{PFev3}\\
\dot{ {\cal W}} &=& -\Theta\, {\cal W} -\frac{1}{2}\,\kappa \rho\,\Sigma-3\Sigma\,{\cal W},\label{PFev4}
\end{eqnarray}
where $\kappa = 8 \pi G/c^4$. The change of (spatial) volume  ${ V}$ is 
\begin{equation}
\dot{ { V} } = { V} \Theta, \label{PFev5}
\end{equation}
and the spatial curvature ${\cal R}$ follows from the ``Hamiltonian'' constraint
\begin{equation}
\frac{1}{6} {\cal R}= \frac{1}{3}\,\kappa   \rho + \Sigma^2  - \frac{1}{9} \Theta^2 + \frac{1}{3} \Lambda. \label{hamcon}
\end{equation}
Spatially homogeneous and isotropic FLRW models form a small subset of solutions of the above equations, with

\begin{equation}
 \Sigma \equiv 0, \quad {\cal W} \equiv 0, \quad \Theta \equiv 3 \frac{\dot{a} }{a}, \quad  {\cal R} \equiv 6\frac{ k}{a^2},  \label{flrw}
\end{equation}
where $k$ is a constant, and $a(t)$ is a function of time. Using the equation for $\dot{\Theta}$ and the ``Hamiltonian'' constraint the evolutionary equations can be written in a more familiar form of the Friedmann equations

\begin{eqnarray}
&& 3 \frac{\ddot{a}}{a} = - \frac{1}{2} \kappa \rho  +  \Lambda,  \label{fes1} \\ 
&&  3 \frac{\dot{a}^2}{a^2} = \kappa  \rho  - 3 \frac{k}{a^2}   + \Lambda  \label{fes2}.
\end{eqnarray}
Equations  (\ref{fes2}) is sometimes written in terms of $\Omega$s
\begin{equation}
1 = \Omega_m(t) + \Omega_k(t) + \Omega_\Lambda (t),\label{triplet}
\end{equation}
where 
$\Omega_m(t) = \frac{1}{3} \kappa \frac{\rho}{H^2}$, $\Omega_k(t) = - \frac{k}{H^2 a^2}$, $\Omega_\Lambda (t) = \frac{1}{3} \frac{\Lambda }{H^2}$.
This characterizes the evolution of spatially homogeneous FLRW models. Below it is shown that within a general class of the silent universes, the mean spatial curvature $\Omega_{\cal R}$ (in the FLRW limit  $\Omega_{\cal R} \to \Omega_k$) will evolve from zero towards a negative non-negligible curvature.

\subsection{ The average spatial curvature} 

Within the silent universe, each worldline evolves independently, thus the volume average over a domain ${\cal D}$ of a scalar function $f$ is

\begin{equation}
\av{f}_{\cal D}  = \frac{ \sum_i \, f_i \, V_i} {\sum_i V_i}, \label{volav}
\end{equation}
where the volume of the domain ${\cal D}$ is 
$ V_{\cal D} = \sum_i V_i$, 
and the size of the domain is
$r_{\cal D} = \left( \frac{3}{4 \pi} V_{\cal D} \right)^{1/3}$.
Averaging (\ref{hamcon})

\begin{equation}
\frac{1}{6}  \av{ {\cal R} }_{\cal D} = \frac{1}{3}\,\kappa   \av{ \rho }_{\cal D} + \av{ \Sigma^2}_{\cal D}  - \frac{1}{9} \av{ \Theta^2 }_{\cal D} + \frac{1}{3} \Lambda
\end{equation}
and introducing the domain Hubble parameter $H_{\cal D} =  \frac{1}{3} \av{ \Theta }_{\cal D}$
we can rewrite the above equation (cf. eq. (\ref{triplet}))

\begin{equation}
\Omega_\mathcal{R}^{\cal D} = 1 - \Omega_m^{\cal D} - \Omega_\Lambda^{\cal D} - \Omega_\mathcal{Q}^{\cal D}
\end{equation}
where

\begin{eqnarray}
\Omega_\mathcal{R}^{\cal D} &=& -\frac{ \av{\mathcal{R}}_{\cal D} } { 6 H_{\cal D}^2 }, \nonumber \\
\Omega_m^{\cal D} &=& \frac{8 \pi G}{3 H_{\cal D}^2 } \av{\rho}_{\cal D}, \nonumber \\
\Omega_\Lambda^{\cal D} &=& \frac{\Lambda}{3 H_{\cal D}^2}, \nonumber \\
\Omega_\mathcal{Q}^{\cal D} &=& \frac{1} {H_{\cal D}^2 } \left( \av{ \Sigma^2}_{\cal D}  + \frac{1}{9} \av{ \Theta^2 }_{\cal D}  - H_{\cal D}^2 \right).
\label{cqar}
\end{eqnarray}
The above is often referred to as the {\em cosmic quartet} \citep{2008GReGr..40..467B}, and $\Omega_{\cal Q}$ is the {\em kinematic backreaction} \citep{Buchert:1999er} (to be precise, $\Omega_{\cal Q}$ is the dimensionless parameter of kinematic backreaction ${\cal Q}_{\cal D}$, which is defined as ${\cal Q}_{\cal D} = - 6 H^2_{\cal D} \Omega_{\cal Q} = \frac{2}{3}  \av{ \Theta^2}_{\cal D} - 6 H^2_{\cal D}- 2 \av{ \sigma^2}_{\cal D}$, where $2 \sigma^2 = \sigma_{ab} \sigma^{ab} = 6 \Sigma^2$).

The FLRW limit follows from eqs. (\ref{flrw})

\begin{align}
& \Omega_\mathcal{R}^{\cal D} \to \Omega_k = -\frac{k}{H^2 a^2}, \nonumber \\
& \Omega_m^{\cal D} \to \Omega_m = \frac{8 \pi G}{3 H^2 } \, \rho, \nonumber \\
& \Omega_\Lambda^{\cal D} \to \Omega_\Lambda =  \frac{1}{3 H^2} \, \Lambda, \nonumber \\
& \Omega_\mathcal{Q}^{\cal D} \to 0 .  \nonumber 
\end{align}

\subsection{Setting up a Monte Carlo simulation of silent universes} 

The evolution of the universe is studied by performing a 
Monte Carlo simulation that is based on tracing the evolution of 
$2 \times 10^6$ worldlines.
Each worldline is evolved by solving eqs. (\ref{PFev1})--(\ref{PFev4}).
This is done with the code \textsl{simsilun}\footnote{\texttt{https://bitbucket.org/bolejko/simsilun}} \citep{simsulun.paper}.
The initial conditions are set as perturbations around the $\Lambda$CDM background -- it is assumed that the early universe is well approximated by the $\Lambda$CDM model. The $\Lambda$CDM model is defined by $\Omega_m = 0.308$, $\Omega_\Lambda = 0.692$, and $H_0 = 67.81$ km s$^{-1}$ Mpc$^{-1}$ \citep{2016A&A...594A..13P}. The initial instant is set to be an instant, which corresponds to $z_i = 200$ in the $\Lambda$CDM. This is to ensure smallness of initial perturbations, and to minimize the effect of pressure, which for $z<200$ is below a percent level. 

The initial density perturbation $\delta_i$, for each worldline, is drawn 
from a Gaussian distribution with a standard deviation $\sigma_i = 0.0094$.
This, ensures that the present-day standard deviation of the density contrast evaluated within spheres of radius  $8 h^{-1}$ Mpc (which is a standard  definition of the cosmological parameter $\sigma_8$) is $\sigma_8 =  0.82$ \citep{2016A&A...594A..13P}.

Since the Monte Carlo simulation considered here, consists of $2 \times 10^6$ worldlines,
thus $2 \times 10^6$ initial density contrasts $\delta_i$ are generated.
These are then used to set up the initial conditions for $\rho_i$,
$\Theta_i$, $\Sigma_i$, and ${\cal W}_i$ that are needed to solve eqs. (\ref{PFev1})--(\ref{PFev4}).
The initial conditions are evaluated based on the initial $\delta_i$ \citep{simsulun.paper}.
\begin{align}
& \rho_i = \rho_{\Lambda CDM} + \Delta \rho = \rho_{\Lambda CDM} \, ( 1 + \delta_i), \nonumber \\
& \Theta_i = \Theta_{\Lambda CDM} + \Delta \Theta =  \Theta_{\Lambda CDM} \, ( 1 - \frac{1}{3}  \, \delta_i), \nonumber \\
& \Sigma_i = -\frac{1}{3} \, \Delta \Theta, \nonumber \\
& {\cal W}_i =   - \frac{1}{6} \, \Delta \rho, \label{inits}
\end{align}
where $8 \pi G \, \rho_{\Lambda CDM} = \Omega_m 3 H_0^2 (1+z_i)^3 $, and the value of the cosmological constant is  $\Lambda = \Omega_\Lambda 3 H_0^2$.

In addition, the volume around each worldline is calculated using eq. (\ref{PFev5}).
The initial volume $V_i$ follows from $V_i = M_i/\rho_i$, where it is assumed that each worldline has the same mass of $M_i = 1.6 \times 10^{14} M_{\odot}$. This value has been chosen for two reasons: 1) it is a mass of a small cluster or a small void, and 2) with this initial condition, the present day radius of an average cell (i.e. volume around each worldline) is $8 h^{-1}$ Mpc. 
Thus, the total mass and the present-day size of a simulated universe are of order $10^{20} M_{\odot}$ and $10^9$ pc, respectively.

\begin{figure}
\begin{center}
\includegraphics[scale=0.95]{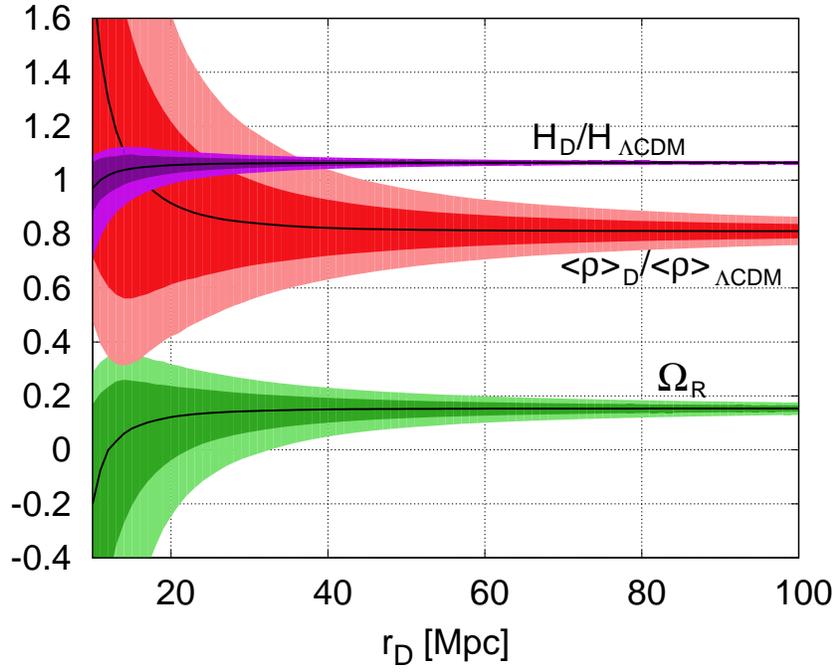}
\end{center}
\caption{Volume averages at the present instant. Darker colors present the 68\% scatter (i.e. 68\% of values of averaging over random domains of size $r_{\cal D}$ fall within this interval) and lighter colors present 95\% (i.e. 95\% of values of averaging over random domains of size $r_{\cal D}$ fall within this interval). Beyond 100 Mpc the cosmic variance is negligible. }
\label{fig1}
\end{figure}

\section{ Emergence of spatial curvature} \label{emcur}

The evolution of each worldline, within the studied Monte Carlo simulation,
is evaluated from eqs. (\ref{PFev1})--(\ref{PFev4}) using the code
\textsl{simsilun} \citep{simsulun.paper}.
This allows for evaluating such quantities as density, expansion rate, and shear, from which 
using (\ref{hamcon}) the spatial curvature can be estimated. 
Using  eq. (\ref{PFev5}) the evolution of the volume is evaluated, and then
using eq. (\ref{volav}) the volume averages of these quantities are calculated.
The resulted average density $\av{\rho}_{\cal D}$,  expansion rate $H_{\cal D}$, and spatial curvature $\Omega_\mathcal{R}^{\cal D}$ are presented in Fig. \ref{fig1}. These averages are evaluated by taking random domains of radius $r_{\cal D}$. 
For small values of $r_{\cal D}$, the density distribution is highly asymmetric with a long tail towards highly dense regions, which is a feature 
of a log-normal distribution observed in the galaxy surveys \citep{2004LRR.....7....8L}. At $r_{\cal D} = 8 h^{-1}$ Mpc $ \approx 11.8$  Mpc, the standard deviation of the density field is $\sigma_8 =  0.82$.

\begin{figure}
\begin{center}
\includegraphics[scale=0.95]{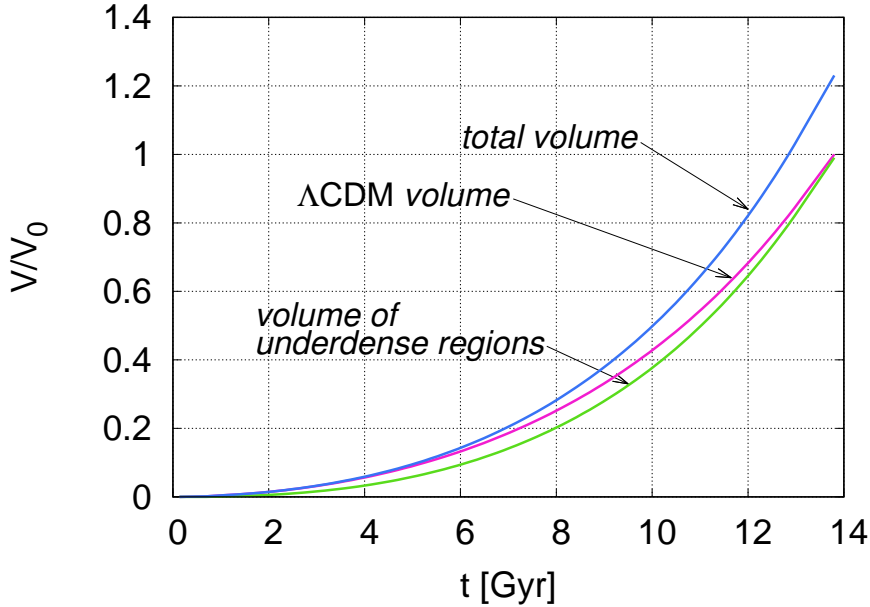}
\end{center}
\caption{Evolution of the volume normilaised by the present day volume of the $\Lambda$CDM model $V_0$: the total volume of the simulated Styrofoam universe (blue), the $\Lambda$CDM model (magenta), the volume of the regions with $\rho < 0.75 \rho_{\Lambda CDM}$ (green).}
\label{fig2}
\end{figure}

The most striking feature is that the mean does not coincide with the $\Lambda$CDM model.
For $r_{\cal D} > 100$ Mpc, where the cosmic variance is negligible, the means are: $\av{\rho} = 0.811 \, \rho_{\Lambda CDM}$, $\av{H} = 1.065 \, H_0$, and $\Omega_{\cal R} = 0.153$.
The reason why these means are not consistent with the $\Lambda$CDM model is presented in Fig. \ref{fig2}. Figure \ref{fig2} shows the evolution of volume of the simulated universe
(i.e. the volume occupied by $2 \times 10^6$ worldlines that constitute the Monte Carlo simulation).
 As seen the volume is larger than in the 
$\Lambda$CDM model (the volume of the $\Lambda$CDM model was calculated using the same setup, i.e.
 $2 \times 10^6$ worldlines  but with $\delta_i = 0$, and it has been checked that the evolution indeed follows eqs. (\ref{fes1}) and (\ref{fes2}) with (\ref{flrw})).
Since the mass is conserved and volume is larger, thus the average density is lower compared to the $\Lambda$CDM model.
Another important feature presented in Fig. \ref{fig2} is that the underdense regions, defined as regions with $\rho < 0.75 \rho_{\Lambda CDM}$, occupy at the present instant approximately $80\%$ of the total volume. 

This phenomenon is of a nonlinear origin.
As long as the perturbations are small and within the linear regime, their contribution to $\rho$ and $\Theta$ is negligibly small and as seen from eqs. (\ref{PFev1})
and  (\ref{PFev2}) all regions in the universe evolve in the same way. Once the evolution becomes nonlinear, both shear $\Sigma^2$ and density $\rho$, efficiently slow down the expansion rate of the overdense regions. Consequently they expand more slowly than the underdense regions. Subsequently, as follows from eq. (\ref{PFev5}), underdense regions occupy more volume than overdense regions, which is depicted in Fig. \ref{fig2}.

\begin{figure}
\begin{center}
\includegraphics[scale=0.95]{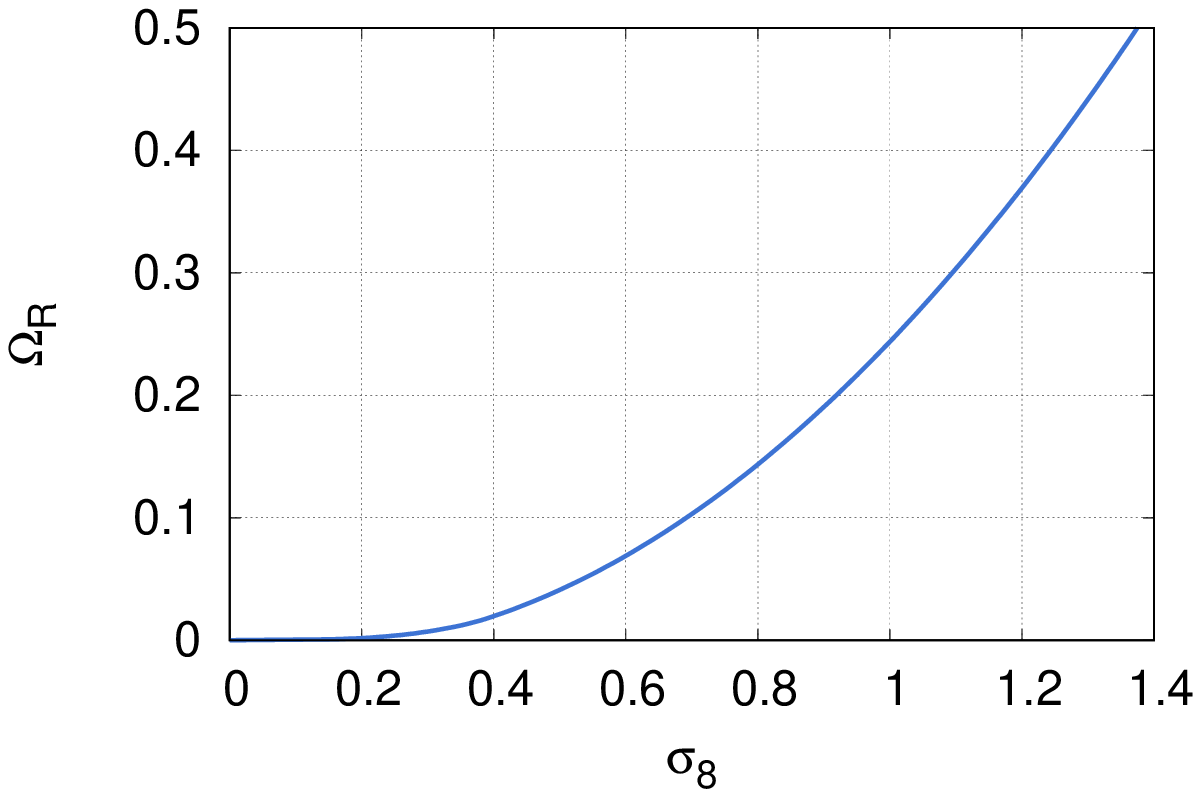}
\end{center}
\caption{The role of nonlinear effects. The linear regime corresponds to small variance of the density field ($\sigma_8 <0.1$) and at these scales the spatial curvature  
averaged over a domain of volume $V = (2150\,{\rm Mpc})^3$ vanishes. With the increase of inhomogeneity, the system enters the nonlinear evolution quicker, leading also to higher values of the density variance, and resulting with large values of the global mean spatial curvature.}
\label{fig3}
\end{figure}

The phenomenon of breaking the symmetry between the evolution of the underdense and overdense regions leads to emerging  negative spatial curvature of the universe.
Tiny initial density perturbations,  as follows from eq. (\ref{hamcon}), are also 
associated with curvature perturbations (negative for underdense and positive for overdense regions).
In the course of evolution these tiny initial perturbations grow. The growth of the spatial curvature is also present in the FLRW model. As seen from eq. (\ref{triplet}),  $\Omega_k = -k/\dot{a}^2$. So if initially $k\ne 0$ then  $\Omega_k$ is also non-zero and evolve as $\dot{a}^{-2}$.
As long as the growth is linear the global average of the spatial curvature is zero --  negative contribution from underdense regions is compensated by positive curvature of overdense regions.
However, once the symmetry of evolution is broken by nonlinear evolution, the mean spatial curvature deviates from zero. This process is depicted in Fig. \ref{fig3}, which shows the relation between the present-day density variance measured with the parameter $\sigma_8$ and the present-day value of $\Omega_{\cal R}$. As described above, apart from the mass $M_i$ (which is kept constant and the same for all worldlines), the only free parameter is the variance of the initial density field $\sigma_i$. The higher the initial variance $\sigma_i$, the larger the amplitude of initial perturbations, which eventually leads to a larger variance of the present-day density field (measured by $\sigma_8$).  The larger the amplitude of initial perturbations, the quicker 
the evolution becomes nonlinear. Consequently the larger $\sigma_i$ and $\sigma_8$, the larger the present-day spatial curvature $\Omega_{\cal R}$.  On the other hand, 
if the initial variance $\sigma_i$ is very small (and so $\sigma_8$)
the evolution of structures stays within the linear regime and does not enter the nonlinear stage. Subsequently, as seen in Fig. \ref{fig3}, for $\sigma_8 < 0.1$ the universe in still within the linear regime and stays spatially flat, confirming that the emergence of the spatial curvature is a feature of nonlinear evolution.

\begin{figure}
\begin{center}
\includegraphics[scale=0.95]{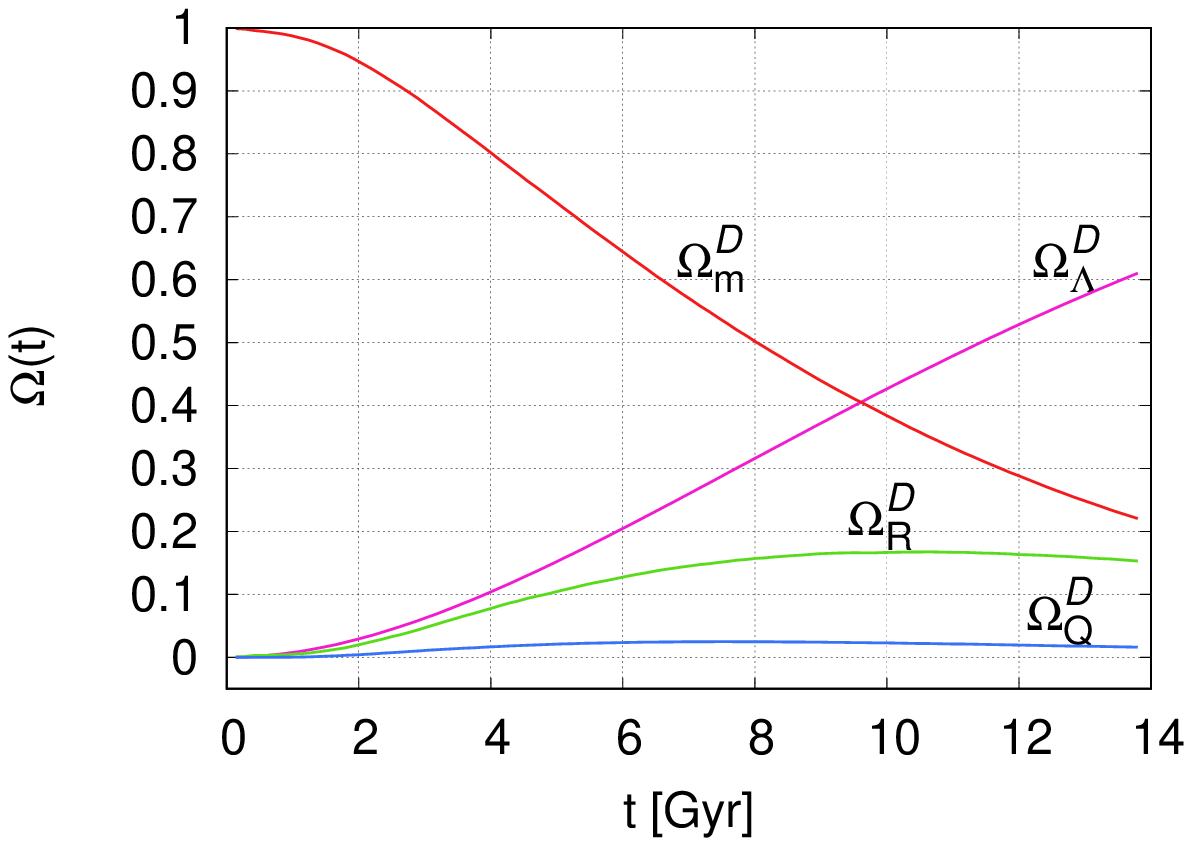}
\end{center}
\caption{Cosmic quartet and its evolution (cf. eqs. (\ref{cqar})), with averages evaluated over a domain of mass $M = 3.2 \times 10^{20} M_{\odot}$.  In the $\Lambda$CDM model, the present day values are $\Omega_m = 0.308$, $\Omega_\Lambda = 0.692$, and $\Omega_k = 0$ (in the FLRW limit $\Omega_{\cal R} \to \Omega_k$ and $\Omega_{\cal Q} = 0$). }
\label{fig4}
\end{figure}

The evolution of the spatial curvature $ \Omega_{\cal R}$ and other cosmic constituents, as defined by eqs. (\ref{cqar}), is presented in Fig. \ref{fig4}. At the initial stages the system is 
dominated by matter  but with time, the  cosmological constant starts to dominate.
In addition, when the evolution enters the nonlinear stage, the spatial curvature emerges, but then around 10 Gyr later it starts to decrease, which is related to the accelerated expansion caused by the cosmological constant. As above, this can be easily understood using the FLRW analogy; in the FLRW regime $\Omega_k \sim \dot{a}^{-2}$, so when the evolution start to accelerate, $\dot{a}$ starts to increase and consequently $\Omega_k$ decreases. This is also known as the cosmological ``no-hair'' conjecture, which states that the  dark energy dominated universe asymptotically approaches a homogeneous and isotropic de Sitter state \citep{1991AnP...503..518P}, and as a results both $ \Omega_{\cal R}$ and $ \Omega_{\cal Q}$ after initial increase, asymptotically approach zero.

\begin{figure}
\begin{center}
\includegraphics[scale=0.95]{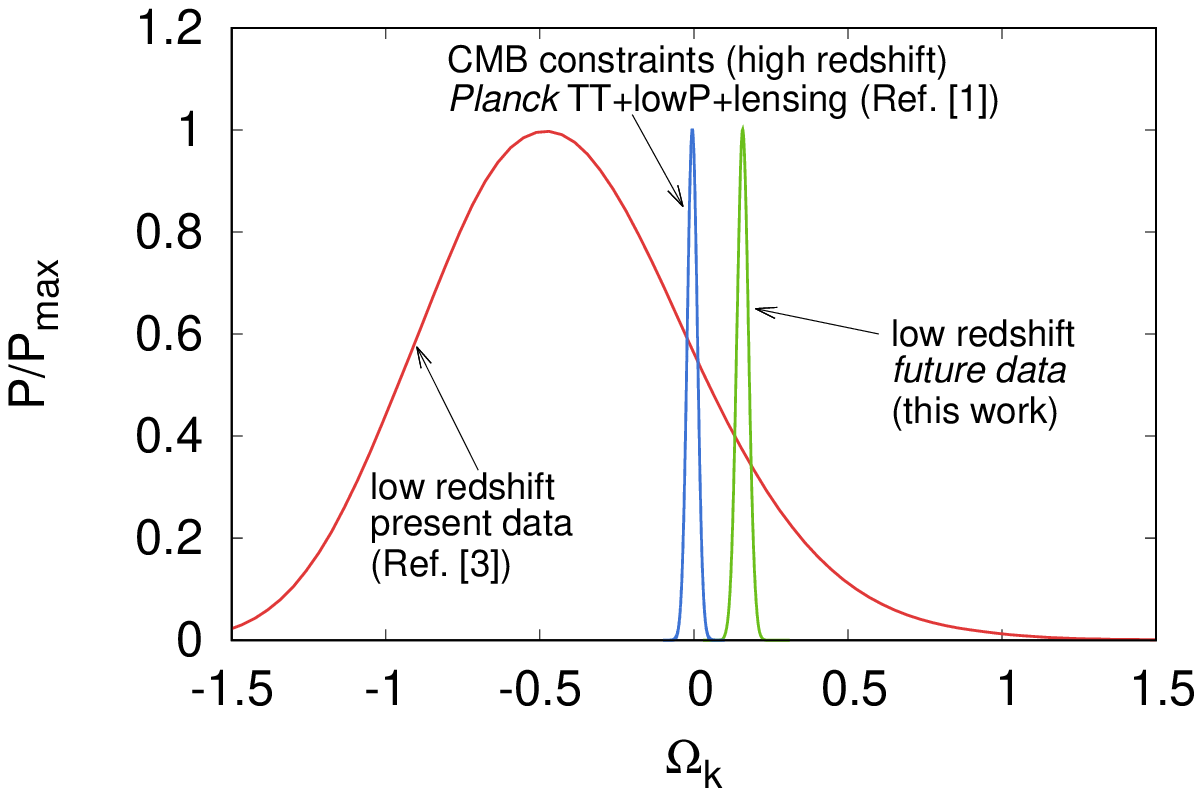}
\end{center}
\caption{Constraints and predictions on the spatial curvature. The high redshift constraints (from CMB) are compared with the low-redshift constraints (based on the method described by \citet{2015PhRvL.115j1301R}). The current data are based on the Supplemental Material of \citet{2015PhRvL.115j1301R} and the future predictions are based on $2 \cdot 10^4$ simulated data ($10^4$ supernova data and $10^4$ lensing data). 
In near future (5--7 years) we will be able to directly measure the curvature of the low-redshift universe and compare it to the spatial flatness of the high-redshift universe.}
\label{fig5}
\end{figure}

\section{Conclusions} \label{conclusion}

This paper examined the emergence of the negative spatial curvature within a general class of silent universes. The analysis was based on the Monte Carlo simulation of $2\times 10^6$ silent universes with the total mass of $3.2 \times 10^{20} M_\odot$. 
The initial conditions were set as perturbations around the  $\Lambda$CDM model
at the instant corresponding to $z=200$. The analysis showed that 
once the evolution becomes nonlinear, the symmetry between the evolution of underdense and overdense regions is broken, and underdense regions start to occupy most of the volume of the Universe, reaching $80\%$ at the present instant (Fig. \ref{fig2}). 
In the linear regime, the spatial curvature averages to $0$ (see Fig. \ref{fig3}) but once the evolution is nonlinear the mean evolution deviates from $0$ and evolves 
towards $\Omega_{\cal R}^{\cal D}  = 0.15$ at the present instant (see Fig. \ref{fig4}).

Another important results  is that the current expansion rate is
$\av{H}_{\cal D} = 72.2$ km s$^{-1}$ Mpc$^{-1}$, which is by $6\%$ higher compared to the  $\Lambda$CDM model of the early universe (see Fig. \ref{fig1}). This difference is a natural consequence of the universe dominated by voids, and could be an obvious solution to the widely debated discrepancy between the 
Hubble constant inferred from low-redshift observations 
$H_0 = 73.24 \pm 1.74$ km s$^{-1}$ Mpc$^{-1}$ \citep{2016ApJ...826...56R} and the one derived from the high-redhsift data (CMB) $H_0 = 67.81 \pm 0.92$ km s$^{-1}$ Mpc$^{-1}$ \citep{2016A&A...594A..13P}.

The model discussed here, although quite general, has some limitations. The silent universes do not have pressure waves or gradients. As a result each worldline evolves independently. At scales below 2 Mpc where particle fluxes, multiple eigenvalues of the shear field, and rotation 
are important, the model looses its accuracy \citep{2002PhRvD..66l4015E}. 
There are some indications that the small scales effects ($< 1$ Mpc) such as viralization  \citep{2013JCAP...10..043R,2017arXiv170606179R} and the environmental dependent clock rates \citep{2009PhRvD..80l3512W} may enhance the discussed effects, but it seems that the fully relativistic modeling of these scales will only be achieved with numerical relativity \citep{Bentivegna:2015flc,Mertens:2015ttp,2017PhRvD..95f4028M}.

Therefore, the phenomenon of emerging curvature, presented here, should be considered as a theoretical speculation, and more work is needed to established the full magnitude to this phenomenon.
What is encouraging, though, is that this phenomenon follows naturally from the nonlinear cosmic dynamics, and has a potential to explain some observational tensions such as a conflict between the low-redshift and high-redshift measurements of $H_0$.   
It is also worth pointing out that the phenomenon of emerging curvature will 
soon be directly testable with observational data.
As shown by \citet{2015PhRvL.115j1301R}, using the data from surveys such as Euclid, and DES and LSST, we  will be able to construct a Gpc-scale `cosmic triangle' (with sides: observer-lens, observer-source, lens-source) and directly measure the curvature of the low-redshift  universe ($z<1$).
By comparing low-redshift spatial curvature  with the constraints obtained from high-redshift data (CMB), we will be able to test if the low-redshift and high-redshift constraints are different, which would empirically prove the phenomenon of the emerging spatial curvature. 
Such a procedure, has been presented in Fig. \ref{fig5}. Applying the procedure discussed and presented by \citet{2015PhRvL.115j1301R}
to the Monte Carlo simulation discuses in Sec. \ref{emcur},
the expected constraints are $\Omega_k = 0.15^{+0.04}_{-0.03}$ ($95\%$ confidence level). 
These constraints are based on simulated $10^4$ lensed galaxies (expected from Euclid survey) and $10^4$ supernova (expected from DES and LSST surveys), and are presented in  Fig. \ref{fig5}.
For comparison, Fig. \ref{fig5} also shows constraints from present low redshift-data (supernova and lensing data obtained from the Supplemental Material of \cite{2015PhRvL.115j1301R}) and high-redshift data ({\em Planck} TT+LowP+lensing). Thus, as seen from Fig. \ref{fig5}, in the next 5--7 years the phenomenon of emerging spatial curvature will be directly testable.

\section*{Acknowledgement}
This work was supported by the Australian Research Council through the Future Fellowship FT140101270. 

\bibliographystyle{mnras}
\bibliography{ms} 

\end{document}